\documentclass[9pt,journal]{IEEEtran}
\newtheorem{theorem}{Theorem}

\newtheorem{lemma}{Lemma}
\newtheorem{definition}{Definition}
\newtheorem{remark}{Remark}

\newtheorem{assumption}{Assumption}

\usepackage{bm}
\usepackage{graphicx}
\usepackage{subfigure}
\usepackage{array}
\usepackage{amsmath}
\usepackage{mathrsfs} 
\usepackage{amssymb}
\usepackage{cite}
\usepackage{enumerate} 
\usepackage{float}
\usepackage{amsfonts} 
\usepackage{epsfig} 
\usepackage{epstopdf}
\usepackage[dvipsnames]{xcolor}
\usepackage{tikz}\newcommand*{\circled}[1]{\lower.7ex\hbox{\tikz\draw (0pt, 0pt)%
		circle (.4em) node {\makebox[1em][c]{\small #1}};}}
\usepackage{color}

\usepackage[colorlinks,
linkcolor=blue,
anchorcolor=blue,
citecolor=blue]{hyperref}

\begin{document}
	
\title{A Novel Edge Laplacian-based Approach for Adaptive Formation Control of Uncertain Multi-agent Systems with Unified Relative Error Performance}
\author{Kun Li, Kai Zhao, \IEEEmembership{Member, IEEE}, Yongduan Song, \IEEEmembership{Fellow, IEEE}, and Lihua Xie, \IEEEmembership{Fellow, IEEE}
\thanks{K. Li, K. Zhao, and Y. D. Song are with School of Automation, Chongqing University, Chongqing 400044 China (e-mail: likun/zhaokai/ydsong@cqu.edu.cn).}
\thanks{L. H. Xie is with School of Electrical and Electronic Engineering, Nanyang Technological University, 639798 Singapore (e-mail: elhxie@ntu.edu.sg).}}
	
\maketitle

\begin{abstract}
	For most existing prescribed performance formation control methods, performance requirements are not directly imposed on the relative states between agents but on the consensus error, which lacks a clear physical interpretation of their solution. In this paper, we propose a novel adaptive prescribed performance formation control strategy, capable of guaranteeing prescribed performance on the relative errors, for uncertain high-order multi-agent systems under a class of directed graphs. Due to the consideration of performance constraints for relative errors, a coupled nonlinear interaction term that contains global graphic information among agents is involved in the error dynamics, leading to a fully distributed control design more difficult and challenging. Here by proposing a series of nonlinear mappings and utilizing the edge Laplacian along with Lyapunov stability theory, the presented formation control scheme exhibits the following appealing features when compared to existing results: 1) different performance requirements can be guaranteed in a unified way by solely tuning the design parameters a priori, without the need for control redesign and stability reanalysis under the proposed fixed control protocol, making the design more user-friendly and the implementation less demanding; 2) the complex and burdensome verification process for the initial constraint, often encountered in existing prescribed performance controls, is completely obviated if the performance requirements are global; and 3) nonlinear interaction is completely decoupled and the asymptotic stability of the formation manifold is ensured via using the adaptive parameter estimate technique. Finally, simulations of various performance behaviors are performed to show the efficiency of the theoretical results.
\end{abstract}
	
\begin{IEEEkeywords}
	Adaptive control, edge Laplacian, formation control, prescribed performance.
\end{IEEEkeywords}
	
\interdisplaylinepenalty=2500

\section{INTRODUCTION}
Cooperative control of multi-agent systems (MASs) has attracted considerable attention in recent decades due to its broad application potential. Representative problems include consensus \cite{olfati2004consensus}, formation control \cite{ren2007distributed}, and converge control \cite{zhong2011distributed}, to name a few. Wherein, formation control, the goal of which is to coordinate each agent using local information so that the whole team forms a prescribed spatial geometric pattern \cite{korea2015}, has been widely studied as it is applicable to various practical tasks, such as transporting loads, exploring resources, and environmental monitoring \cite{underwater}.

In addition to the classical formation control objectives, MASs may also need to meet some additional steady and transient behaviors due to performance requirements or safety constraints in physical, e.g., completion time, maximum allowable overshoot, tolerable range of tracking accuracy, maintaining connectivity, and collision avoidance \cite{wei2022adaptive}. If we ignore such performance requirements, it may lead to the failure of cooperative formation tasks. Let us provide two specific examples to illustrate the necessity of performance and safety constraints in formation control. Consider the multi-robot hand used in neurosurgery, a large overshoot or slow response can present a risk to life. Therefore, the controller used in the robotic hand should not only have a desired steady-state response but also a desired transient response. In addition, for a platoon of vehicles to cooperatively transport a common load, the distances between any two vehicles in the team must be neither too small nor too large, as this can lead to collisions between the vehicles or a breakdown of the link. 

Current efforts on such issues are primarily based on the artificial potential function \cite{sun2018}, barrier Lyapunov function \cite{restrepo2022robust}, funnel control (FC) \cite{min2023low,verginis2023asymptotic,lee2022synchronization}, and prescribed performance bound (PPB) control \cite{bechlioulis2016decentralized,verginis2017robust,cheng2022fixed,wang2015prescribed}. Notably, only the FC and PPB methodologies can be used to guarantee the prescribed performance behavior of the MASs in terms of overshoot, convergence rate and steady-state error. {Yet, it is essential to emphasize that, whereas such performance specifications are obtained, these FC and PPB-based works only take into account the performance of the sum of relative errors between agent $i$ and all its neighbors (consensus errors) and not the relative errors between agent $i$ and each neighbor, which does not reflect the inter-agent physical restrictions.} Taking the cooperative formation control as an example, it is indeed more practical to specify the performance constraints on the relative position of agents, so that the interest of collision avoidance and connectivity maintenance can be tackled. On the other hand, from the user's perspective, it is more intuitive to define constraint functions based on the relative errors between agents, as the constraint requirements are normally shaped by the environmental boundaries.

In recent years, some typical results on the performance characteristics of relative errors between agents have been developed. In \cite{karayiannidis2012multi} and \cite{macellari2016multi}, consensus protocols with certain prescribed performance guarantees were proposed for single- and double-integrator dynamics, respectively. In \cite{bechlioulis2014robust}, a decentralized formation control protocol with connectivity maintenance was presented for first-order nonlinear MASs. In \cite{chen2020leader}, the relative position-based formation control for leader-follower MASs with prescribed performances was studied. In \cite{stamouli2022robust}, a distributed dynamic average consensus algorithm was proposed for MASs under undirected communication topologies, so that the consensus with predefined and arbitrary convergence rate is guaranteed. Recently, a novel connectivity preservation and collision-free formation control algorithm for double-integrator dynamics was proposed in \cite{huang2024prescribed}. 

{By reviewing the above performance-constrained works \cite{karayiannidis2012multi,macellari2016multi,bechlioulis2014robust,chen2020leader,huang2024prescribed,stamouli2022robust}, we can conclude that 1) the results only consider the single- or double-integrator dynamics under undirected graphs, which is impractical in the real applications since many engineering systems suffer from the high-order dynamics and modeling uncertainties; and 2) \textcolor[rgb]{0,0,0}{there exists an initial constraint on the relative errors in the algorithm implementation, which implies that the aforementioned methods only guarantee the semi-global results. To achieve the performance requirements, it is necessary to look for suitable initial values of system states and then to verify the initial constraint. However, due to the complexity and large-scale network structure of MASs, the verification process will become more and more demanding and grow explosively as the number of \textcolor[rgb]{0,0,0}{agents} increases (the detailed discussion can be found in Section \ref{asymmetric}).} Moreover, when the considered system is restarted or the target formation is changed, we have to check whether the original design parameters are applicable to the new initial data. If such a condition is not met, the corresponding control is no longer available, leading to the inflexibility of the control method. These limitations pose numerous obstacles to their application, which in turn motivates this work.
	
Motivated by the above observations, we will investigate the relative-error-performance formation problem for uncertain high-order MASs under directed graphs. As mentioned earlier, the difficulties in solving these problems are mainly in two aspects, 1) \textcolor[rgb]{0,0,0}{how to design a fully distributed formation protocol with prescribed performance guarantees for uncertain high-order MASs under directed graphs}, and 2) how to ensure different performance behaviors between an agent and its neighbors within a fixed control framework according to its own environmental constraints. To address the above challenges, we present a unified prescribed performance solution for the formation control problem of uncertain high-order MASs. Compared with the existing results, the main contributions of this paper can be listed as follows:
	\begin{enumerate}[1)]
		\item \textbf{Plug and Play Function:} The proposed framework features that different performance requirements for the relative errors can be guaranteed by tuning the design parameters a priori and are completely decoupled from the system model, topology graph, and control gain selection. \textcolor[rgb]{0,0,0}{It provides a new formulation that achieves different prescribed performance for relative position errors of uncertain high-order nonlinear MASs in a unified way, without changing the control structure.} Especially, the complex and explosive verification process for the initial constraints in existing prescribed performance controls is completely obviated if the performance requirement is in the global form;
		\item \textbf{User-Oriented Feature:} Compared with the consensus error-based PPB schemes \cite{bechlioulis2016decentralized,verginis2017robust,cheng2022fixed,wang2015prescribed}, the proposed solution enables user to establish performance constraint boundaries directly based on environmental constraints, making it more favorable in practical applications. This physics-based performance control design approach provides a physical interpretation of the proposed approach; and
		\item \textbf{Broad Applicability:} We propose an adaptive formation control scheme for uncertain high-order nonlinear MASs under a class of directed and undirected graphs. In this respect, existing methods in \cite{karayiannidis2012multi,macellari2016multi, bechlioulis2014robust,chen2020leader,stamouli2022robust,huang2024prescribed} that consider simple dynamics under undirected graphs are contained as special cases of our work.
	\end{enumerate}
	
	\emph{Notations:} In this paper, $||\cdot||$ is the Euclidean norm. The sets of real, positive real, and $n$-dimensional Euclidean space are $\mathbb{R}$, $\mathbb{R}^+$, and $\mathbb{R}^n$, respectively. Let $\textbf{1}_n$ be unit column vector with $n$-dimensional. $X^\top$ denotes the transpose of a matrix $X$. For $X_i \in \mathbb{R}^{m \times n}, i=1,\dots,q$, we denote $\textrm{diag}[X_i]= \textrm{blockdiag}\{X_1,\dots, X_q\}\in \mathbb{R}^{qm \times qn}$. For a set of numbers $V$, $|V|$ is its cardinality. $\textrm{Null}(A)$ stands for the null space of matrix $A$. $\lambda_{\min}(A)$ and $\lambda_{\max}(A)$ denote the minimum and maximum eigenvalue of matrix $A$ respectively. 
	
	\section{PRELIMINARIES}\label{II}
	\subsection{Graph \& Edge Laplacian}
	We consider a team of $N$ agents whose information exchange is described by a {\it directed} graph $\mathcal{G}\triangleq(\mathcal{V,E})$, where $\mathcal{V}=\{1,...,N\}$ is the set of nodes and $\mathcal{E}\subseteq \mathcal{V}\times \mathcal{V}$ with cardinality $m$ is the set of edges. An edge $(i, j) \in \mathcal{E}$ indicates that agent $j$ can obtain information from agent $i$, but not necessarily vice versa. A {\it directed path} in a directed graph is a sequence of edges, for example, $(i_1, i_2), (i_2, i_3), \dots, (i_{N-1}, i_N)$, and a {\it directed cycle} is a directed path where the initial and final nodes are the same. If $(i, j) \in \mathcal{E}$ implies $(j, i) \in \mathcal{E}$, the graph is said to be {\it undirected}. The adjacency matrix $\mathcal{A}=[a_{ij}] \in \mathbb{R}^{N \times N}$ with $a_{ij}=1$, if $(i,j)\in \mathcal{E}$, and $a_{ij}=0$, otherwise. We denote the set of neighbors for agent $i$ as $\mathcal{N}_i=\{j\in \mathcal{V}| (i,j)\in \mathcal{E}\}$. The degree matrix is $\triangle=\text{diag}[\triangle_i]\in \mathbb{R}^{N \times N}$ with $\triangle_i=\sum_{{j\in\mathcal{N}_i}} a_{ij}$ being the sum of the $i$th row in $\mathcal{A}$. Then, the {\it graph Laplacian} matrix of $\mathcal{G}$ is defined as $L=\triangle - \mathcal{A}$. 
	
	To ensure the specified performance for relative errors in the formation control, we choose an alternative graph-theoretic tool, i.e., the edge Laplacian\cite{zelazo2007agreement}. Now we recall this concept as follows.
	
	Label all edges of the digraph $\mathcal{G}$ with $e_1,e_2,\dots,e_m$. Suppose edge $(i, j)$ in $\mathcal{G}$ corresponds to the $k$th directed edge, where $k\in I= \{1, \dots, m\}$. The incidence matrix $E(\mathcal{G})\in \mathbb{R}^{N\times m}$ for the digraph is a $\{0,\pm1\}$-matrix with rows and columns indexed by nodes and edges of $\mathcal{G}$, respectively, such that
	\begin{equation*}
	[E(\mathcal{G})]_{ik}:=\begin{cases}
	+1, &\textrm{if $i$ is the initial node of edge $e_k$},\\ -1, &\textrm{if $i$ is the terminal node of edge $e_k$}, \\ 0, &\textrm{otherwise.}
	\end{cases}
	\end{equation*}
	From the definition of the incidence matrix, it follows that the null space of its transpose, $\textrm{Null}(E(\mathcal{G})^\top)$, contains $\textrm{span}\{\textbf{1}_N\}$.
	Moreover, for the purpose of analysis, it is necessary to partition the incidence matrix as $E(\mathcal{G})=E_{\odot}(\mathcal{G})+E_{\otimes}(\mathcal{G}),$ where $E_{\odot}(\mathcal{G})\in \mathbb{R}^{N\times m}$ corresponds to the so-called {\it in-incidence} matrix, whose elements are defined as
	\begin{equation*}
	[E_{\odot}]_{ik}:=\begin{cases}
	-1, &\textrm{if $i$ is the terminal node of edge $e_k$}, \\ 0, &\textrm{otherwise.}
	\end{cases}
	\end{equation*}
	and $E_{\otimes}(\mathcal{G})\in \mathbb{R}^{N\times m}$ corresponds to the so-called {\it out-incidence} matrix, whose elements are defined as
	\begin{equation*}
	[E_{\otimes}]_{ik}:=\begin{cases}
	+1, &\textrm{if $i$ is the initial node of edge $e_k$},\\ 0, &\textrm{otherwise.}
	\end{cases}
	\end{equation*}
	According to the definitions of incidence matrix and in-incidence matrix, the {edge Laplacian} matrix $L_e\in \mathbb{R}^{m\times m}$ and graph Laplacian $L\in \mathbb{R}^{N\times N}$ of a digraph $\mathcal{G}$ can be defined in terms of the incidence and in-incidence matrices as $L_e=E^\top E_{\odot}$ and $L=E_{\odot}E^\top$. For an undirected graph, $L_e=E^\top E$ and $L=EE^\top$, respectively. 	{Moreover, using an appropriate permutation of the edge ordering, the incidence matrix is expressed as $E=[E_t\quad E_c]$, where $E_t\in \mathbb{R}^{N\times(N-1)}$ denotes the full-column-rank incidence matrix corresponding to an arbitrary spanning tree $\mathcal{G}_t\subset \mathcal{G}$ and $E_c\in \mathbb{R}^{N\times(m-N+1)}$ represents the incidence matrix corresponding to the remaining edges not contained in $\mathcal{G}_t$. Defining $R:=[I_{N-1}\quad T]$, where $T:=\left(E^\top_tE_t\right)^{-1}E^\top_tE_c$, one obtains representation of the incidence matrix of the graph, given by $E=E_tR$. Furthermore, inspired by \cite{restrepo2021edge}, the following facts will play a crucial role in the controller design and stability analysis presented later in this paper.
	\begin{lemma}\cite{restrepo2021edge}\label{restrepo}
	For a digraph that is a spanning tree, it holds that $L^s_e=\frac{1}{2}\left(E^\top E_{\odot}+E^\top_{\odot}E\right)$ is positive definite; For a digraph that is a directed cycle, it holds that $E^\top_tE_t$ is positive definite.
	\end{lemma}}

	\subsection{Problem Statement}\label{III}
	In this paper, we consider a group of $N$ nonlinear agents that can be described as:
	\begin{align}\label{dynamic}
	\begin{cases}
	\dot{x}_{i,q}= x_{i,q+1},~q=1,\dots,n-1, \\
	\dot{x}_{i,n}= u_i+\varphi^\top_i(\overline{x}_{i})\theta_i,~i=1,\dots,N,
	\end{cases}
	\end{align}
	where $x_{i,q}\in \mathbb{R}$ and $u_i\in \mathbb{R}$ are the system state and control input of agent $i$, respectively, where we only consider the one dimension case in the theoretical analysis without loss of generality. The discussions in this work can be generalized to higher dimensions cases with Kronecker product. $\overline{x}_i:=[x^\top_{i,1},x^\top_{i,2},\cdots,x^\top_{i,n}]^\top\in \mathbb{R}^{n}$, $\varphi_i: \mathbb{R}^{n} \rightarrow \mathbb{R}^{\nu_i \times n}$ is a known smooth nonlinear function, and $\theta_i\in \mathbb{R}^{\nu_i}$ represents an unknown constant vector, where $\nu_i\in\mathbb{R}^+$. 
	
	
	Let $e_k:=x_{i,1}-x_{j,1}\in\mathbb{R}$ be the relative position between a pair of neighboring agents $i$ and $j$ for $(i,j)\in\mathcal{E}$, and let $e^d_{k}\in\mathbb{R}$ denote the desired relative position for $e_k$. The  target relative position-based formation is assumed to be admissible and is described as $\mathcal{F}:=\{e|e_k=e^d_k, k\in\mathcal{I}\}$, where $e =E^\top x_1\in\mathbb{R}^m$ denotes the stack vector of the relative states, and $x_1=[x_{1,1},x_{2,1},\dots,x_{N,1}]^\top\in \mathbb{R}^N$ denotes the stack vector of the first-order components. Then we define the \emph{formation errors} as $\tilde{e}_k=e_k-e^d_k\in \mathbb{R}$ for $k\in\mathcal{I}$. It can be checked that the target formation is achieved if $e_k = 0$. The control objective in this paper is formally stated as follows.
	
	
	
	{\it{\textbf{Control Objective:}}} By using the local available sensing measurements from the agent itself and its neighbor agents, we aim to design a distributed adaptive formation controller $u_i$ for \eqref{dynamic} such that: 
	\begin{enumerate}
		\item[$\mathcal{O}_1$:] All signals in the closed-loop system are bounded, and the desired formation, described by the relative offsets $\{e^d_k\}_{k\in\mathcal{I}}$, is achieved and maintained; and 
		\item[$\mathcal{O}_2$:] Multiple prescribed performance requirements for the relative position errors $\{\tilde{e}_k\}_{k\in\mathcal{I}}$ are guaranteed in a unified control framework without control redesign.
	\end{enumerate}
	
	To this end, the following assumption is imposed.
	\begin{assumption}\label{a1}
		The sensing graph should either be a connected undirected graph or a directed graph that is a directed spanning tree or a directed cycle.
	\end{assumption}
	
	
	\section{System and Performance Transformation}
	In this section, we formulate the formation performance control problem by constructing a unified performance function. Motivated in part by a previous work \cite{zhao2021adaptive}, we propose a more comprehensive unified performance function to enhance the applicability of the algorithm.
	
	%
	%
	
	\subsection{Unified Performance Function}
	\begin{definition}\label{de1}
		Let $\mathcal{P} : (-\iota, \iota) \rightarrow (-\infty,\infty)$ be a unified performance function for $\iota\in \mathbb{R}^+$, which satisfies the following conditions.
		\begin{enumerate}
			\item $\lim_{y\rightarrow\iota}\mathcal{P}(y) = \infty$, $\lim_{y\rightarrow -\iota}\mathcal{P}(y) = -\infty$, and $\mathcal{P}(0)=0$;
			\item  The derivative of $\mathcal{P}$ is lower bounded by a positive constant $\underline{q}$ and tends to infinity as $y\rightarrow \pm \iota$; and
			\item $\mathcal{P}$ is continuously differentiable and $\mathcal{P}(-y) = -\mathcal{P}(y)$.
		\end{enumerate}
	\end{definition}
	Obviously, immediate examples of such $\mathcal{P}(y)$ include (but are not limited to) the following functions: $
	\mathcal{P}(y)=\frac{y}{\sqrt{1-y^2}}$ with $\iota=1$ and $ \mathcal{P}(y)=\tan(y)$ with $\iota=\frac{\pi}{2}$. 
	Utilizing the unified performance function definition, the second objective $\mathcal{O}_2$ can be mathematically stated as:
	\begin{align}
	\mathcal{P}_k(-\underline{\delta}_k\beta_{k}(t))<\tilde{e}_{k}(t)<\mathcal{P}_k(\overline{\delta}_k\beta_{k}(t))~\textrm{for}~k\in I, \label{e-bound}
	\end{align}
	where $0<\underline{\delta}_{k}, \overline{\delta}_k\le 1$ are user-chosen parameters. Additionally, $\beta_{k}(t)$ is a continuously decaying performance positive function with bounded derivative and a strictly positive limit as $t\rightarrow \infty$, which is formulated as $\beta_{k}:=(\beta_{k0}-\beta_{kf})\exp(-\lambda_k t)+\beta_{kf}$
	with $0<\beta_{kf}<\beta_{k0}\le \iota$ and $\lambda_k\in \mathbb{R}^{+}$.
	
	It can be readily seen that by choosing different design parameters $\underline{\delta}_{k}$, $\overline{\delta}_{k}$, and $\beta_{k0}$, such a performance objective as defined in (\ref{e-bound}) covers different cases of interest, including {\it lower/upper bounded one-sided} performance constraints, {\it asymmetric} performance constraints, and {\it global} performance constraints, for details see below.
	\begin{enumerate}
		\item If $\underline{\delta}_{k}{\beta_{k0}}=\overline{\delta}_{k}{\beta_{k0}}=\iota$, then (\ref{e-bound}) can represent global performance constraints in the form of $-\infty <\tilde{e}_k(0)<\infty$; 
		\item If $\overline{\delta}_{k}{\beta_{k0}}=\iota$ and $0<\underline{\delta}_{k}<1$ (resp. $\underline{\delta}_{k}{\beta_{k0}}=\iota$ and $0<\overline{\delta}_{k}<1$), then (\ref{e-bound}) can represent {lower bounded one-sided} performance constraints in the form of $-\underline{\epsilon}_{k}<\tilde{e}_k(0)<\infty$ (resp. {upper bounded one-sided} performance constraints in the form of $-\infty<\tilde{e}_k(0)<\overline{\epsilon}_{k}$), where $\underline{\epsilon}_{k}=\mathcal{P}_k(-\underline{\delta}_k\beta_{k0})$ and $\overline{\epsilon}_{k}=\mathcal{P}(\overline{\delta}_k\beta_{k0})$ are some positive constants; and
		\item If $0<\underline{\delta}_{k}{\beta_{k0}}<\iota$ and $0<\overline{\delta}_{k}{\beta_{k0}}<\iota$, then (\ref{e-bound}) can also represent asymmetric performance constraints in the form of $-\underline{\epsilon}_{k}<\tilde{e}_k(0)<\overline{\epsilon}_{k}$. 
	\end{enumerate}
	
	Overall, the designer has the flexibility to choose various design parameters (i.e., $\underline{\delta}_k, \overline{\delta}_k$, and $\beta_{k0}$) based on practical requirements. It is interesting to underscore that in the case of asymmetric performance constraints, the existing relative error PPB-based results \cite{karayiannidis2012multi,macellari2016multi, bechlioulis2014robust,chen2020leader,stamouli2022robust,huang2024prescribed} are relevant. Their algorithms are special cases of the algorithm developed in this paper.

	\subsection{Transformed Error Dynamic Model}
	To deal with the  constraints in (\ref{e-bound}), we will adopt the following nonlinear mapping:
	\begin{align}
	s_{k} &:=\frac{\zeta_{k}(t)}{(\underline{\delta}_{k}+\zeta_{k}(t))(\overline{\delta}_{k}-\zeta_{k}(t))}, \label{s} \\
	\zeta_{k} &:=\frac{\eta_k(t)}{\beta_{k}(t)}, ~\eta_k :=\mathcal{P}^{-1}_k(\tilde{e}_k(t)) \label{zeta}
	\end{align}
	with $\zeta_{k}(t)$ denoting the modulated error and $\eta_{k}(t)$ being the normalized error. Furthermore, the following lemma can be easily derived from the aforementioned transformation.
	
	\begin{lemma}\label{Property 1}
		If $\zeta_{k}(0)$ satisfies $-\underline{\delta}_{k}<\zeta_{k}(0)<\overline{\delta}_{k}$ and $s_{k}(t)$ is bounded for $\forall t\ge0$, then there exist some positive constraints $\underline{\delta}_{1k}$ and $\overline{\delta}_{1k}$ so that $-\underline{\delta}_{k}<-\underline{\delta}_{1k}\le \zeta_{k}(t) \le \overline{\delta}_{1k} < \overline{\delta}_{k}$. Furthermore, (\ref{e-bound}) can be guaranteed.
	\end{lemma} 
	
	{\it{\textbf{Proof.}}} The proof of this lemma is similar to \cite{zhao2021adaptive}, therefore it is omitted herein. $\hfill\blacksquare$
	
	It is thus obvious that if an adaptive control law is designed such that $s_k(t)$ is bounded over $t\in [0, \infty)$, the target $\mathcal{O}_2$ for the tracking performance is achieved according to the Lemma \ref{Property 1}. 
	To facilitate the descriptions in control design and stability analysis, the unified performance function $\mathcal{P}(y)$ in this work is chosen as $\mathcal{P}(y)={y}/{\sqrt{1-y^2}}$. Upon using the expression of $s_k$ as shown in (\ref{s}), with the definitions of $\eta_k$ and $\beta_{k}$ in (\ref{zeta}), it is seen that
	\begin{align}\label{ds}
	\dot{s}_{k}=\frac{\partial s_{k}}{\partial \zeta_{k}}\frac{\textrm{d} \zeta_{k}}{\textrm{d}t}=J_{k}\left(\xi_{k}\dot{\tilde{e}}_{k}-{\dot{\beta}_{k}}\zeta_{k}\right),
	\end{align}
	where $J_{k}=\frac{\mu_{k}}{\beta_{k}}>0$, $\mu_{k}=\frac{\underline{\delta}_{k}\overline{\delta}_{k}+{\zeta}^2_{k}}{(\underline{\delta}_{k}+\zeta_{k})^2(\overline{\delta}_{k}-\zeta_{k})^2}>0$,  $\xi_{k}=\frac{{1}}{\left(1+\tilde{e}^2_{k}\right)\sqrt{\tilde{e}^2_{k}+1}}>0$
	are available for control design.
	
	Based on the definition of the formation error $\tilde{e}_k$, the transformed error dynamics for system (\ref{ds}) are rewritten in the following compact form
	\begin{align}\label{dS}
	\dot{S}=&J\left( \xi \dot{\tilde{e}}+\dot{\beta}\zeta\right)=WE^\top x_2+D,
	\end{align}
	where $S=[s_1,\dots,s_m]^\top\in \mathbb{R}^m$, $J=\textrm{diag}[J_k]\in \mathbb{R}^{m\times m}$, $\xi=\textrm{diag}[\xi_k]\in \mathbb{R}^{m\times m}$, $\dot{\beta}=\textrm{diag}[-{\dot{\beta}_k}] $, $\zeta=[\zeta_1,\dots,\zeta_m]^\top$, $W=J \xi$, and $D=J\dot{\beta}\zeta$. Furthermore, given that each argument of  $W$ is positive, it follows that the diagonal matrix $W$ is positive definite.
	
	By replacing the equation of $\dot{x}_{i,1}$ in (\ref{dynamic}) with $\dot{s}_k$ in (\ref{ds}), the performance constraint objective is transformed to the stabilization problem of the following strict-feedback-like system:
	\begin{equation}\label{strict system}
	\begin{cases}
	\dot{S}=WE^\top x_2+D,   \\
	\dot{x}_{q}=x_{q+1},~q=2,\dots,n-1,  \\
	\dot{x}_{n}=u+\varphi^\top\theta,
	\end{cases}
	\end{equation}
	where $x_q=[x_{1,q}, x_{2,q},\dots, x_{N,q}]^\top$, $u=[u_{1}, u_{2},\dots, u_{N}]^\top$, $\varphi=\textrm{diag}[\varphi^\top_1, \varphi^\top_2, \dots, \varphi^\top_N]$, and $\theta=[\theta_{1}, \theta_{2},\dots, \theta_{N}]^\top$.
	

	\section{Controller Design and Stability Analysis}
	
	\subsection{Controller Design}
	In this {subsection}, based on the adaptive backstepping technique \cite{krstic1995nonlinear} we propose a novel adaptive formation design approach {by} utilizing the edge Laplacian. Firstly, we introduce the following change of coordinates:
	\begin{align}
	z_{1}= &~S, \\
	z_{q}= &~x_{q}- \alpha_{q-1} \label{zq}, q=2,\dots,n,
	\end{align}
	where $z_1=[s_1, \dots, s_{m}]^\top$ and $z_q=[z_{1,q}, \dots,z_{N,q}]^\top$ are intermediate variables, and $\alpha_{q-1}=[\alpha_{1,q-1}, \dots, \alpha_{N,q-1}]^\top$ is the virtual controller, respectively. The design procedure contains $n$ steps. 
	
	\textbf{ Step 1:} From (\ref{dynamic}), (\ref{dS}) and (\ref{zq}), the derivative of $z_1$ is
	\begin{align}\label{dz1}
	\dot{z}_1= WE^\top(z_2+\alpha_{1})+D,
	\end{align}
	then the derivative of $V_1=\frac{1}{2}z^\top_1z_1$ along  (\ref{dz1}) yields,
	\begin{equation}\label{dv1}
	\dot{V}_1=z^\top_1\left[WE^\top(z_2+\alpha_{1})+D\right].
	\end{equation}
	
	\begin{remark}
		\textcolor[rgb]{0,0,0}{We pause here to emphasize that, due to the consideration of relative error performance behaviors, the coupled nonlinear interaction term $WE^\top$ and the additional coupling term $J\dot{\beta}\zeta$ have been incorporated into the derivative of $V_1$ as presented in (\ref{dv1}). This introduces an obstacle in designing the virtual controller $\alpha_1$ based on local information to ensure the controllability of the subsystem dynamics (\ref{dz1}). The main challenges arise from the following three aspects: 
		\begin{enumerate}
			\item The transformed error dynamics, as described in (\ref{ds}), are significantly different from the conventional PPC control methodologies\cite{bechlioulis2016decentralized,verginis2017robust,cheng2022fixed,wang2015prescribed}. This distinction stems from the inclusion of a nonlinear term $\xi_k$, which is coupled with the formation errors, leading to difficulties in ensuring the controllability of subsystems (\ref{dz1});
			\item The term $WE^\top$ represents a non-square matrix that contains global topological information, making the direct use of its inverse impractical for controller design; and
			\item The third challenge arises from the fact that in the setup for directed graphs, the Laplacian matrix $L$ is non-symmetric and positive semi-definite. Thus, it is challenging to ensure the controllability of the subsystem dynamics (\ref{dz1}) using Laplacian matrix-based design methods.	
		\end{enumerate}}
	\end{remark}
	
	To handle the above challenges, we perform the controller design of $\alpha_{1}$ using the edge Laplacian technique. The virtual controller $\alpha_{1}$ is designed as
	\begin{equation}\label{alpha1}
	\alpha_{1}=-c_1E_{\odot} W z_1, ~~c_1>0.
	\end{equation}
	
	By substituting (\ref{alpha1}) into (\ref{dv1}), the derivative of $V_1$ is given by
	\begin{align}\label{dv11}
	\dot{V}_1=-c_1z^\top_1WE^\top E_{\odot} W z_1+z^\top_1WE^\top z_2+z^\top_1D.
	\end{align}
	
	\begin{remark}
		We highlight that each component of $\alpha_{1}$ depends only on local information since $E_{\odot}$ represents the incoming edges on each node, i.e., the information available to each agent as defined by the directed graph. Moreover, equation (\ref{dv11}) is applicable to both directed spanning-tree, directed-cycle and undirected graph topologies. In the following, we proceed by separately examining each of these topologies.
	\end{remark}
	
	\emph{Case 1 (Directed spanning tree):} In this case, we have $\mathcal{G}_t=\mathcal{G}$. Therefore, $E_t=E$, $R=I_{N-1}$, and $E_{\odot t}=E_{\odot}$, where $E_{\odot t}$ is the in-incidence matrix of the spanning tree. It is not difficult to obtain $z_{1t}=z_1$ since $R=I_{N-1}$. Then we have
	\begin{align}\label{dv1alpha}
	\dot{V}_1&=-c_1z^\top_{1t}WE^\top_t E_{\odot t} W^\top z_{1t}+z^\top_{1t}WE^\top_t z_2+z^\top_{1t}D\nonumber\\
	&=-c_1z^\top_{1t}WL^s_e W^\top z_{1t}+z^\top_{1t}WE^\top_t z_2+z^\top_{1t}D\nonumber\\
	&\le -{c_1\lambda_{\min}(L^s_e)}\Vert W z_{1t}\Vert^2+z^\top_{1t}WE^\top_t z_2+z^\top_{1t}D,
	\end{align}
	where the third inequality follows from the fact that $L^s_e$ is positive definite according to Lemma \ref{restrepo}. 
	
	Then after using $D=J\dot{\beta}\zeta$ in (\ref{dS}) and applying Young’s inequality to last two term in the right-hand side of (\ref{dv1alpha}), we obtain
	\begin{align}
	z^\top_{1t} WE^\top_t z_2 &\le \frac{\vartheta}{2}\lambda_{\max}(E^\top_t E_t)\Vert Wz_{1t}\Vert^2+\frac{1}{2\vartheta}\|z_2\|^2,  \label{yang1}\\
	z^\top_{1t} J \dot{\beta} \zeta &\le   \frac{\epsilon}{2}\Vert Jz_{1t}\Vert^2 +\frac{1}{2\epsilon}\zeta^\top \dot{\beta}  \dot{\beta} \zeta,  \label{yang2}
	\end{align}
	where $\vartheta$ and $\epsilon$ are positive constants.
	
	Now, given $c_1>0$, let $\vartheta>0$ be such that $c'_1:=\left[c_1\lambda_{\min}(L^s_e)-\frac{1}{2}\vartheta\lambda_{\max}(E^\top_t E_t)\right]\lambda_{\min}(W^2)$ and $c'_2=c_2-\frac{1}{2\vartheta}$ with $c_2>0$ are positive, we bound the (\ref{dv1alpha}) as follows:
	\begin{align}\label{dV1spanning}
	\dot{V}_1&\le -{c'_1}\Vert z_{1t}\Vert^2+\frac{1}{2\vartheta}\|z_2\|^2+\frac{\epsilon}{2}z^\top_{1t}J J z_{1t}+\frac{1}{2\epsilon}\zeta^\top\dot{\beta}\dot{\beta}\zeta \nonumber \\
	& \le  -{c''_1}\Vert z_{1t}\Vert^2+\frac{1}{2\vartheta}\|z_2\|^2+\frac{1}{2\epsilon}\zeta^\top\dot{\beta}\dot{\beta}\zeta,
	\end{align}
	where $c''_1:=c'_1-\frac{\epsilon}{2}\lambda_{\max}(J^2)$ is positive for a sufficiently small $\epsilon>0$.
	
	\emph{Case 2 (Directed cycle):} In this case, we have $E^\top E_{\odot}+E^\top_{\odot}E=E^\top E$. By using this fact, we obtain 
	\begin{align}\label{dv1cycle}
	\dot{V}_1&=-\frac{1}{2}c_1z^\top_1WE^\top EW z_1+z^\top_1WE^\top z_2+z^\top_1D\nonumber\\
	&=-\frac{1}{2}c_1z^\top_{1t}RWR^\top E^\top_t E_t R W R^\top z_{1t}+z^\top_{1}W E^\top z_2+z^\top_1D\nonumber\\
	&\le -{\frac{1}{2}c_1\lambda_{\min}(E^\top_tE_t)}\Vert RW R^\top  z_{1t}\Vert^2+z^\top_{1t}RWR^\top E^\top_t z_2\nonumber\\
	&\quad +z^\top_1J\dot{\beta}\zeta,
	\end{align}
	where the third inequality follows from the fact that $E^\top_t E_{t}$ is positive definite according to Lemma \ref{restrepo}. 
	
	Now, given $c_1>0$, let $\vartheta>0$ be such that $c'_1:=\frac{1}{2}\left[c_1\lambda_{\min}(E^\top_tE_t)-\vartheta\lambda_{\max}(E^\top_t E_t)\right]\lambda_{\min}(RWR^\top RWR^\top )$ and $c'_2=c_2-\frac{1}{2\vartheta}$ with $c_2>0$ are positive. Note that $RW R^\top RW R^\top$ and $RJ^2R^\top$ are positive definite due to $R$ is full row-rank and $W$ and $J$ are positive definite. Then, similar to the case 1, by applying Young’s inequality to last two term in the right-hand side of (\ref{dv1cycle}), we obtain
	\begin{align}\label{dV1cycle}
	\dot{V}_1&\le -{c'_1}\Vert z_{1t}\Vert^2+\frac{1}{2\vartheta}\|z_2\|^2+\frac{\epsilon}{2}z^\top_{1t}RJ JR^\top z_{1t}+\frac{1}{2\epsilon}\zeta^\top\dot{\beta}\dot{\beta}\zeta \nonumber \\
	& \le  -{c''_1}\Vert z_{1t}\Vert^2+\frac{1}{2\vartheta}\|z_2\|^2+\frac{1}{2\epsilon}\zeta^\top\dot{\beta}\dot{\beta}\zeta,
	\end{align}
	where $c''_1:=c'_1-\frac{\epsilon}{2}\lambda_{\max}(RJ^2R^\top)$ is positive for a sufficiently small $\epsilon>0$.

	\emph{Case 3 (Undirected graph):} In this setup, each agent can not only receive information from its neighbors but also broadcast its own information. Therefore, the virtual controller can be designed as follows:
	\begin{equation}\label{alpha1undirected}
	\alpha_{1}=-c_1E W z_1,
	\end{equation}
	where $E$ represents the incidence matrix of the oriented undirected graph. 
	
    By substituting (\ref{alpha1undirected}) into (\ref{dv1}), the derivative of $V_1$ is given by
	\begin{align}\label{dv1undirected}
	\dot{V}_1&=-c_1z^\top_1WE^\top EW z_1+z^\top_1WE^\top z_2+z^\top_1D\nonumber\\
	&=-c_1z^\top_{1t}RWR^\top E^\top_t E_t R W R^\top z_{1t}+z^\top_{1}W E^\top z_2+z^\top_1D\nonumber\\
	&\le -{c_1\lambda_{\min}(E^\top_tE_t)}\Vert RW R^\top  z_{1t}\Vert^2+z^\top_{1t}RWR^\top E^\top_t z_2\nonumber\\
	&\quad +z^\top_1J\dot{\beta}\zeta,
	\end{align}
	where $E^\top_tE_t$ represents the edge Laplacian of a spanning tree $\mathcal{G}_t\subset \mathcal{G}$ and is symmetric positive definite. 
    
    It is seen from (\ref{dv1undirected}) that $\dot{V}_1$ has a similar expression. Therefore, by employing the same analysis as in Case 2, we can easily obtain:
	\begin{align}\label{dV1undirected}
	\dot{V}_1 & \le  -{c''_1}\Vert z_{1t}\Vert^2+\frac{1}{2\vartheta}\|z_2\|^2+\frac{1}{2\epsilon}\zeta^\top\dot{\beta}\dot{\beta}\zeta,
	\end{align}
	where $c''_1:=c'_1-\frac{\epsilon}{2}\lambda_{\max}(RJ^2R^\top)>0$ and $c'_1:=\left[c_1\lambda_{\min}(E^\top_tE_t)-\frac{1}{2}\vartheta\lambda_{\max}(E^\top_t E_t)\right]\lambda_{\min}(RWR^\top RWR^\top )>0$.
	
	In summary, we see from (\ref{dV1spanning}), (\ref{dV1cycle}), and (\ref{dV1undirected}) that the expression of $\dot{V}_1$ is same for both the considered directed graph and undirected graph. With this in mind, we proceed with the controller design.

	{\textbf{ Step 2:}}  
	Using (\ref{zq}), the derivative of $z_2$ is computed as
	\begin{align}\label{dz2}
	\dot{z}_2=\dot{x}_2-\dot{\alpha}_1=z_3+\alpha_{2}-\dot{\alpha}_1.
	\end{align}
	
	From (\ref{dv1}) and (\ref{dz2}), the derivative of $V_2=V_1+\frac{1}{2}z^\top_2z_2$ is
	\begin{align}\label{dv2}
	\dot{V}_2 \le & -{c''_1}\Vert z_{1t}\Vert^2+\frac{1}{2\vartheta}\|z_2\|^2+\frac{1}{2\epsilon}\zeta^\top\dot{\beta}\dot{\beta}\zeta \nonumber \\
	&+z^\top_2 \left(z_3+\alpha_{2}-\dot{\alpha}_1\right).
	\end{align}
	
	Therefore, we design $\alpha_{2}$ as
	\begin{equation}\label{alpha2}
	\alpha_{2}=-c_2z_2+\dot{\alpha}_1, ~~c_2>0.
	\end{equation}
	
	By replacing $\alpha_2$ from (\ref{alpha2}) in (\ref{dv2}), one has
	\begin{align}\label{dv22}
	\dot{V}_2 \le -c''_1\Vert z_{1t}\Vert^2-c'_2\Vert z_2\Vert^2+z^\top_2 z_3+\frac{1}{2\epsilon}\zeta^\top\dot{\beta}\dot{\beta}\zeta.
	\end{align}
	{\textbf{ Step $q~(q=3, \dots, n-1)$:}} Define the quadratic function $V_q$ as $V_q=V_{q-1}+\frac{1}{2}z^\top_qz_q$, then its derivative is
	\begin{align}\label{dvq}
	\dot{V}_q \le &  -c''_1\Vert z_{1t}\Vert^2-c'_2\Vert z_2\Vert^2-\sum_{j=3}^{q-1}c_{j}z^\top_{j}z_j +z^\top_{q-1} z_{q} \nonumber \\
	& +\frac{1}{2\epsilon}\zeta^\top\dot{\beta}\dot{\beta}\zeta +z^\top_q \left(z_{q+1}+\alpha_{q}-\dot{\alpha}_{q-1}\right).
	\end{align}
	Following the ideas in the above steps, we design $\alpha_{q}$ as
	\begin{equation}\label{alphaq}
	\alpha_{q}=-c_qz_q-z_{q-1}+\dot{\alpha}_{q-1},~~c_q>0.
	\end{equation}
	Substituting the virtual control $\alpha_{q}$ into (\ref{dvq}), one has
	\begin{align}\label{dvqq}
	\dot{V}_q  \le & -c''_1\Vert z_{1t}\Vert^2-c'_2\Vert z_2\Vert^2-\sum_{j=3}^{q-1}c_{j}z^\top_{j}z_j \nonumber \\
	&  +z^\top_q z_{q+1} +\frac{1}{2\epsilon}\zeta^\top\dot{\beta}\dot{\beta}\zeta.
	\end{align}
	{\textbf{ Step n:}} The derivative of $z_n$ is computed as
	\begin{align}\label{dzn}
	\dot{z}_n=\dot{x}_n-\dot{\alpha}_{n-1}=u+\varphi^\top\theta-\dot{\alpha}_{n-1}.
	\end{align}

	Define the Lyapunov function candidate $V$ as
	\begin{equation}\label{Vfunction}
	V=V_{n-1}+\frac{1}{2}z^\top_nz_n+\frac{1}{2}\tilde{\theta}^\top \Gamma^{-1}\tilde{\theta},
	\end{equation}
	where $\tilde{\theta}=\theta-\hat\theta$ is the estimate error of $\theta$ with $\hat\theta$ being the estimate of $\theta$, and $\Gamma=\textrm{diag}[\Gamma_i]\in \mathbb{R}^{N\times N}$ is a positive definite matrix with $\Gamma_i>0$. Then the derivative of $V$ is
	\begin{align}\label{dvn}
	\dot{V} \le &  -c''_1\Vert z_{1t}\Vert^2-c'_2\Vert z_2\Vert^2-\sum_{j=3}^{n-1}c_{j}z^\top_{j}z_j +z^\top_{n-1} z_n   \nonumber \\
	&+\frac{1}{2\epsilon}\zeta^\top\dot{\beta}\dot{\beta}\zeta +z^\top_n \left(u+\varphi^\top\theta-\dot{\alpha}_{n-1}\right)-\tilde{\theta}^\top \Gamma^{-1}\dot{\hat{\theta}}.
	\end{align}
	The actual controller and the adaptive law are designed as:
	\begin{align}
	u&=-c_nz_n-z_{n-1}+\dot{\alpha}_{n-1}-\varphi^\top\hat{\theta}, ~~c_n>0,\label{u}\\
	\dot{\hat{\theta}}&=\Gamma\varphi z_n. \label{adaptive}
	\end{align}
	By applying the actual control and adaptive law into (\ref{dvn}), we have
	\begin{align}\label{dvnn}
	\dot{V} \le &  -c''_1\Vert z_{1t}\Vert^2-c'_2\Vert z_2\Vert^2-\sum_{j=3}^{n-1}c_{j}z^\top_{j}z_j+\frac{1}{2\epsilon}\zeta^\top\dot{\beta}\dot{\beta}\zeta.
	\end{align}

	
	\subsection{Stability Analysis}
	Here we present our main result in the
	following theorem.
	\begin{theorem}\label{T1}
		Consider the uncertain nonlinear MASs described by (\ref{dynamic}) under Assumption \ref{a1}. If the developed control (\ref{u}) and adaptive law (\ref{adaptive}) are employed, then target formation can be achieved, ensuring the satisfaction of multiple performance requirements, as well as the boundedness of all closed-loop signals at all time.
	\end{theorem}
	
	{\it{\textbf{Proof.}}} 
	We carry out the proof of Theorem \ref{T1} by three phases. Firstly, we address the existence and uniqueness of a maximal solution ${\zeta}_{k}(t)$ over the open set $\Omega_{\zeta_k}:= \{\zeta_k(t): \zeta_{k}(t)\in (-\underline{\delta}_{k}, \overline{\delta}_{k}) \}$ for a time interval $[0, \tau_{\max})$ with $\tau_{\max}$ being a positive constant. Secondly, we prove that all internal signals are bounded and $\zeta_{k}(t)$ remains strictly within a subset of $\Omega_{\zeta_k}$ for $t\in [0, \tau_{\max})$ under the control law (\ref{u}) and the adaptive control law (\ref{adaptive}). Finally, we show that $\tau_{\max}$ can be extended to $\infty$ and subsequently $\zeta_{k}(t) \in (-\underline{\delta}_{k}, \overline{\delta}_{k})$ is ensured over time, thus completing the proof.
	
	
	{\it \textbf{Phase I.}} 
	Note that the modulated error $\zeta_{k}(t)$ is given by $\zeta_k=\eta_k/\beta_k$. Differentiating $\zeta_k$ with respect to time yields:
	\begin{align}\label{dzeta}
	\dot{\zeta}=h_{0}(t,\zeta)=\overline{\beta}\left(\xi E^\top(z_2-c_1E_{\odot} W z_1)+\dot{\beta}\zeta\right),
	\end{align}
	where $\overline{\beta}=\textrm{diag}[\frac{1}{\beta_1}, \dots, \frac{1}{\beta_N}]$. Differentiating ${z}_q$ and $\tilde{\theta}$, employing (\ref{alpha1}), (\ref{alpha2}), (\ref{alphaq}) as well as the fact that $z_{q}= x_{q}- \alpha_{q-1}$, and invoking (\ref{u}), (\ref{adaptive}), we arrive at
	\begin{align}
	\begin{cases}
	\dot{z}_1=h_{1}(t,z_2,\zeta)= WE^\top(z_2-c_1E_{\odot} W z_1)+D,  \\
	\dot{z}_q=h_{q}(t,z_{q+1})=z_{q+1}-c_qz_{q},~q=2,\dots,n-1,  \\
	\dot{z}_n=h_{n}(t,z_{n},\tilde{\theta})=-c_nz_n-z_{n-1}+\varphi^\top\tilde{\theta},  \\
	\dot{\tilde{\theta}}=h_{n+1}(t,z_{n})=-\Gamma \varphi z_n.
	\end{cases}
	\end{align}
	
	Thus, the closed-loop dynamic system of $Z(t)=[\zeta(t), z_1(t), \dots, \\z_n(t),\tilde{\theta}(t)]^\top$ may be written in compact form as:
	\begin{equation}
	\dot{Z}=h(t,Z)=[h^\top_{0}(t,\zeta), h^\top_{1}(t,z_2,\zeta), \cdots, h^\top_{n+1}(t,z_{n})]^\top.
	\end{equation}
	
	Let us also define the open set $\Omega_Z=\Omega_{\zeta} \times \mathbb{R}^{n+1}$, where $\Omega_{\zeta}=\Omega_{\zeta_1}\times \Omega_{\zeta_2}\times \dots \times \Omega_{\zeta_m}$. Note that $\zeta_{k}(0)\in (\underline{\delta}_{k}, \overline{\delta}_{k})$, one observes $Z(0)=[\zeta(0), z_1(0), \dots,z_n(0),\tilde{\theta}(0)]^\top \in \Omega_Z$. Since $h(t,Z)$ is continuous in $t$ and locally Lipschitz in $Z$ over the set $\Omega_{Z}$, the hypotheses of Theorem 54 in \cite{sontag} (see p. 476) are satisfied. Consequently, the $\dot{Z}$-system has a unique and maximal solution over the time interval $[0, \tau_{\max})$, ensuing that $\zeta_k(t)\in \Omega_{\zeta_k}$ for all $t \in [0, \tau_{\max})$, i.e., 
	\begin{equation}\label{zeta-bound}
	\zeta_k(t)\in (-\underline{\delta}_{k}, \overline{\delta}_{k}),~ k\in I, ~\forall t \in [0, \tau_{\max}).
	\end{equation}
	{\it \textbf{Phase II.}} 
	According to (\ref{zeta-bound}), one has
	\begin{equation}
	\|\zeta\|^2<\sum_{k=1}^m\max\{\underline{\delta}^2_{k},\overline{\delta}^2_{k}\}=\rho,
	\end{equation}
	where $\rho$ is a positive constant. Then it follows from (\ref{dvnn}) that
	\begin{equation}\label{dV-end}
	\dot{V} \le  -c''_1\Vert z_{1t}\Vert^2-c'_2\Vert z_2\Vert^2-\sum_{j=3}^{n-1}c_{j}z^\top_{j}z_j+\frac{\rho}{2\epsilon^2} \lambda_{\max}(\dot{\beta} ^2)
	\end{equation}
	for $\forall t \in [0, \tau_{\max})$. Integrating both sides of (\ref{dV-end}) in the interval $[0, \tau_{\max})$ yields that
	\begin{align}
	V(t)+ &\int_{0}^{\tau_{\max}}\left(c''_1z^\top_{1t}z_{1t}+c'_2z^\top_2z_	2+\sum_{j=3}^{n}c_{j}z^\top_{j}z_{j}\right)dt \nonumber \\
	&\le V(0)+\frac{\rho}{2\epsilon^2}\int_{0}^{\tau_{\max}} \lambda_{\max}(\dot{\beta} ^2)dt.
	\end{align}
	
	Given that ${\dot{\beta}_{k}}$ is smooth and $\lim_{t\rightarrow\infty}{\dot{\beta}_{k}}=0$, it follows that $\lim_{t\rightarrow\infty}\lambda_{\max}(\dot{\beta} ^2)=0$, implying that $\int_{0}^{\tau_{\max}} \lambda_{\max}(\dot{\beta} ^2)dt$ is bounded. Based on the definition of $V$ in (\ref{Vfunction}), we establish that $z_{1t}$ and $z_q$ ($q=2,\dots,n$) and $\tilde{\theta}$ are bounded and integrable for all $t \in [0, \tau_{\max})$. Consequently, the parameter estimate $\hat{\theta}$ remains bounded over the interval $[0, \tau_{\max})$. Furthermore, the boundedness of $z_q$ and $\tilde{e}$ ensures that $x_{i,q}$ is also bounded, leading to the boundedness of  $\varphi_i(\overline{x}_i)$ and $\dot{\alpha}_{q}$ ($q=1,\dots,n-1$) within the same interval. As $\beta_{k}$ and $p^*$ are continuously differentiable, $\dot{z}_q$ ($q=1,\dots,n$) is also bounded. Then from (\ref{u}) and (\ref{adaptive}), we conclude that $u$ and $\dot{\hat{\theta}}$ are bounded for all $t \in [0, \tau_{\max}) $. Therefore, all signals in the closed-loop systems are bounded over the interval $[0, \tau_{\max})$.
	Additionally, due to the boundedness of $s_k$, according to the Lemma \ref{Property 1}, we have
	\begin{equation}\label{zeta-localbound}
	-\underline{\delta}_{k}<-\underline{\delta}_{1k}\le \zeta_{k}(t) \le \overline{\delta}_{1k} < \overline{\delta}_{k}, k\in I, t\in [0, \tau_{\max}). 
	\end{equation}
	{\it \textbf{Phase III.}} Note that from (\ref{zeta-localbound}), $\zeta_k(t) \in \Omega'_{\zeta_k}:= \{\zeta_k(t)\in [\underline{\delta}_{1k}, \overline{\delta}_{1k}]\}$ for $\forall t\in [0, \tau_{\max}) $ is a nonempty compact subset of $\Omega_{\zeta_k}$. Hence, assuming $\tau_{\max}\le \infty$ and since $\Omega'_{\zeta_k}\subset \Omega_{\zeta_k}$, Proposition C.3.6 in \cite{sontag} (see p. 481) dictates the existence of a time instant $t' \in [0, \tau_{\max})$ such that $\zeta_{k}(t') \notin \Omega'_{\zeta_k}$, which is a clear contradiction. Therefore, $\tau_{\max}=\infty$. Thus, all closed loop signals remain bounded and moreover $\zeta_{k}\in \Omega'_{\zeta_k}\subset \Omega_{\zeta_k}, \forall t>0$. Based on Lemma \ref{Property 1}, (\ref{e-bound}) is guaranteed for $\forall  t\ge0$. In addition, by applying the Barbalat's lemma, we have $\lim_{t\rightarrow\infty}z_{1t}=0$. Note that $z_{1t}=0$ implies $S=0$ since $R$ is full rank, which further implies $\zeta=0$ and $\eta=0$ according to the nonlinear mapping as shown in (\ref{s}) and (\ref{zeta}), and hence $\tilde{e}_k=0$ for $k\in I$. The proof is completed. $\hfill\blacksquare$
	
	
	\begin{remark}\label{global}
		It is worth noting that although most existing works (e.g., \cite{bechlioulis2016decentralized,verginis2017robust,cheng2022fixed,wang2015prescribed}) can guarantee the asymmetric prescribed performance for the virtual consensus error, it remains unclear that whether the relative error satisfies the performance constraint. Moreover, if the users would like to achieve other different performances, one has to redesign the controller and reanalyze the stability of the closed-loop system, making the control implementation more complex and less flexible. Some efforts (e.g., \cite{karayiannidis2012multi,macellari2016multi, bechlioulis2014robust,chen2020leader,stamouli2022robust,huang2024prescribed}) have been developed to achieve the prescribed performance for the relative errors, but it still only guarantees one special constraint form.
		To solve the above issue, in this paper, we introduce a unified performance function (as defined in Definition \ref{de1}) and a series of nonlinear mappings (as shown in (\ref{s}) and (\ref{zeta})), so that the prescribed performance problem of relative errors is converted into the stabilization problem of a new strict-feedback-like system. Under such a framework, multiple kinds of prescribed performance behaviors can be guaranteed only by solely tuning the design parameters a priori, without the need for control redesign and stability reanalysis, making the design more user-friendly and the implementation less demanding. Especially for the global performance requirement case, the tedious and explosive verification process for the initial constraint in the existing methods is completely eliminated.
	\end{remark}
	
	\begin{remark}
		In this work, due to the considerations of multiple performance constraints for relative errors and the digraph setup, a coupled nonlinear interaction term $WE^\top$ that contains global graphic information among agents and an additional coupling term $J\dot{\beta}\zeta$ are inevitably embedded into the error dynamics (\ref{strict system}), therefore, a mismatched term appearing in (\ref{dv2}) may exist in each step of the backstepping design procedure, which poses an additional challenge for the stability analysis and control design. To overcome these difficulties, inspired by \cite{zelazo2007agreement}, we propose an edge Laplacian-based adaptive control method, which exhibits the following appealing characteristics. Firstly, by skillful designing the first virtual controller $\alpha_{1}$, the nonlinear interconnections term $WE^\top$ can be treated as a quadratic function $W^\top L_eW$ as shown in (\ref{dv11}), then with the aid of the reduced-order representation and the eigenvalue analysis, we utilize the adaptive parameter estimate technique to establish the asymptotic stability of the formation manifold. Secondly, here we need to emphasize that multiple performance behaviors can be  achieved without changing the control structure because the performance index has been completely decoupled from the structure of  the dynamic systems and the underlying graphs. Furthermore, to handle the negative effects caused by the mismatched terms $z^\top_1 WE^\top z_2$ and $z^\top_1 J \dot{\beta} \zeta$, the robust technique is used for further analysis as shown in (\ref{yang1}) and (\ref{yang2}). As such, the stability of the closed-loop system can be ensured via utilizing the Lyapunov stability. 
	\end{remark}

	\section{SIMULATION RESULTS}\label{V}
	To illustrate the effectiveness of the proposed formation controller, we consider a group of 5 mobile robots described by \cite{sun2018} interconnected over the directed spanning-tree/directed cycle/undirected graph showed in Fig. \ref{directed}, moving in the plane (i.e., $d=2$) with the following dynamics:
	\begin{align}
	\begin{cases}
	\dot{x}_{i}(t)=v_i(t),~i=1,\dots,5, \\
	\dot{v}_{i}(t)=u_i(t)-\varphi^\top_i(x_i,v_i)\theta_i, \label{vi}
	\end{cases}
	\end{align}
	where $x_i=[x_{i,1},x_{i,2}]^\top \in \mathbb{R}^2$ and $v_i=[v_{i,1},v_{i,2}]^\top\in \mathbb{R}^2$ are, respectively, the position and velocity of the mobile robots, $\varphi^\top_i(x_i,v_i)\theta_i$ is an unknown friction satisfying $\varphi_i=\textrm{diag}[\tanh(10v_{i,1})-\tanh(100v_{i,1}), \tanh(10v_{i,2})-\tanh(100v_{i,2})]$ and $\theta_i=[0.25,0.25]^\top$.
	
	\begin{figure}[t!]
		\renewcommand{\thefigure}{1}	
		\centering	
		\includegraphics[width=3.3in]{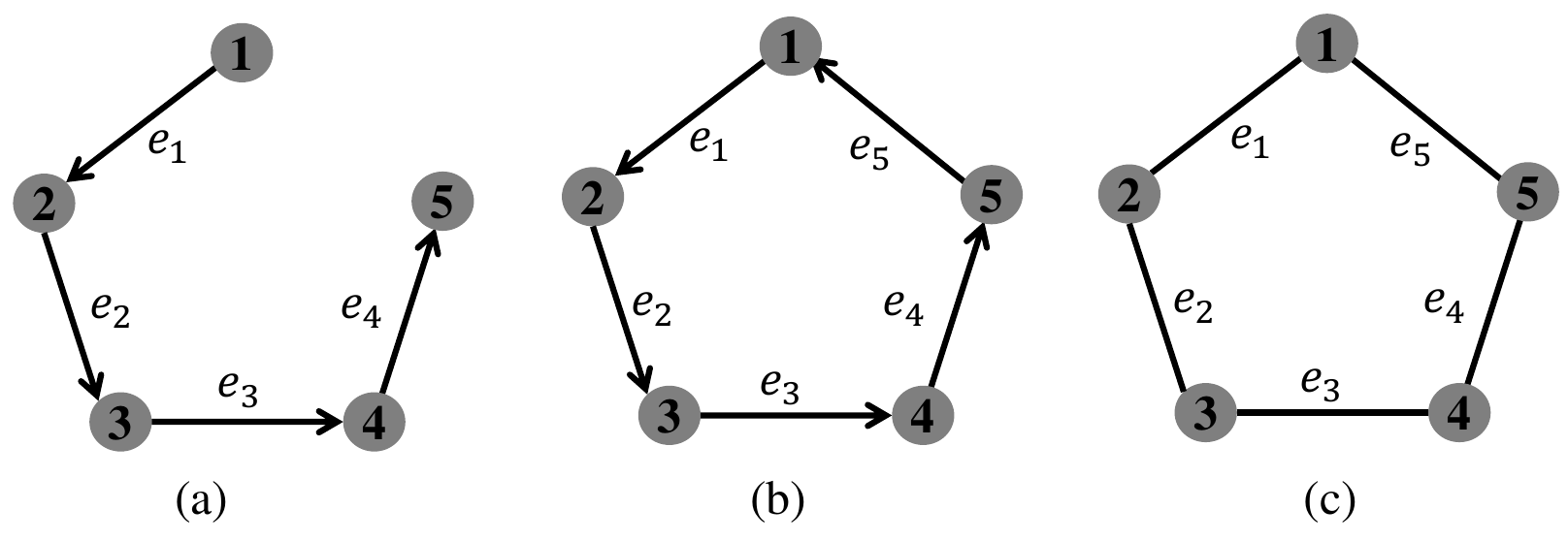}
		\caption{The sensing graph for a group of 5 mobile robots: (a) directed spanning tree; (b) directed cycle; and (c) undirected graph.}\label{directed}
	\end{figure}
	
	\begin{figure*}[t!]
		\renewcommand{\thefigure}{2}
		\centering	
		\includegraphics[width=6.7in]{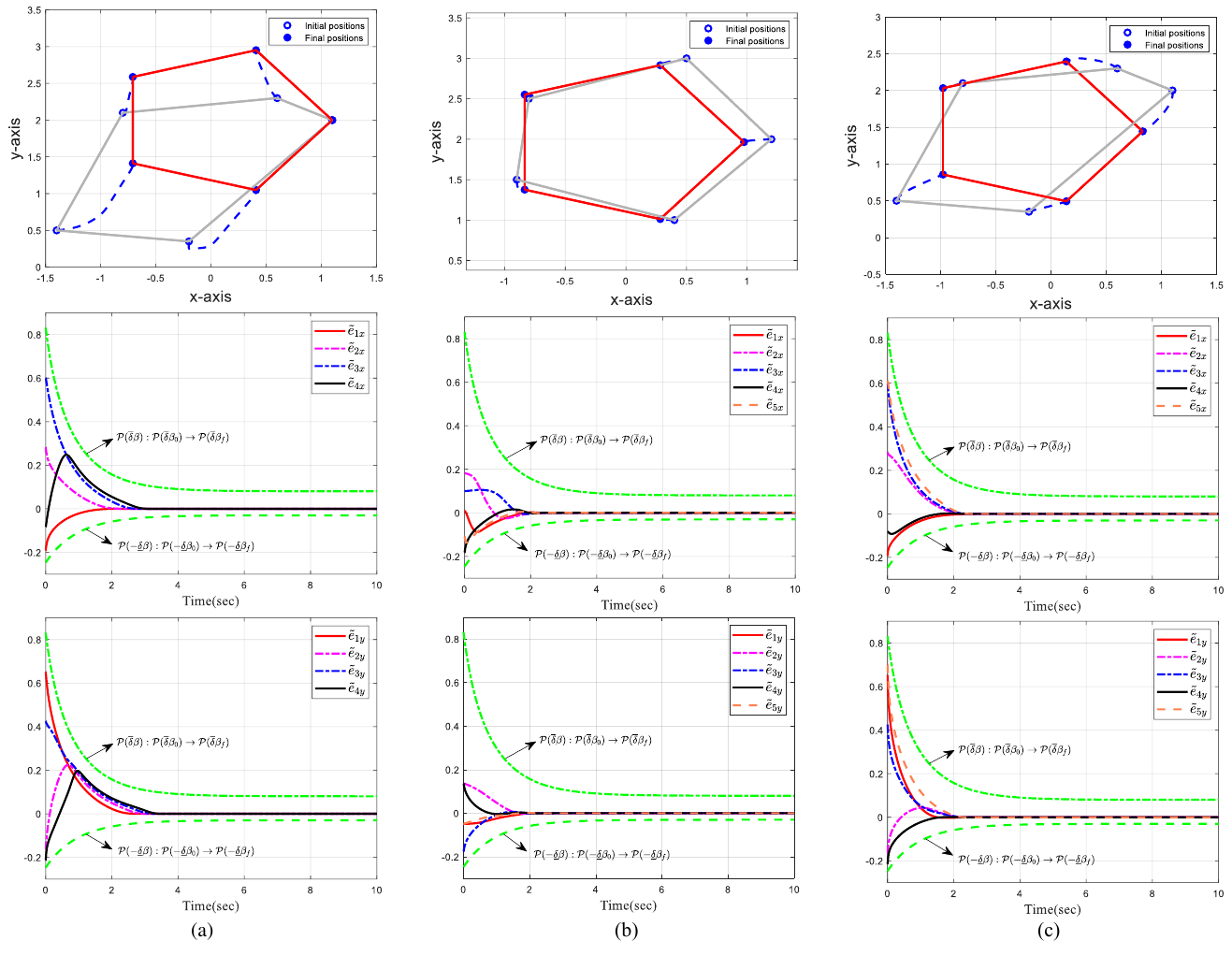}
		\caption{Asymmetric performance behavior. (a) directed spanning tree. (b) directed cycle, and (c) undirected graph cases.}\label{asymmetrycase}
	\end{figure*}
	
	The desired formation corresponds to a pentagon described by $p^d_i(t)=[\cos({2(i-1)\pi}/{5}), \sin({2(i-1)\pi}/{5})]^\top$ for $i=1,\dots,5$, and the desired relative position vector is denoted as $e^d_k=[e^d_{kx}, e^d_{ky}]^\top$ for $k=1,\dots,5$. For Fig. \ref{directed}(a) (directed spanning tree case), $e^d_k$ is set to $e^d_1=p^d_2-p^d_1$, $e^d_2=p^d_3-p^d_2$, $e^d_3=p^d_4-p^d_3$, and $e^d_4=p^d_5-p^d_4$. For Fig. \ref{directed}(b) (directed cycle case),  $e^d_k$ is set to $e^d_1=p^d_2-p^d_1$, $e^d_2=p^d_3-p^d_2$, $e^d_3=p^d_4-p^d_3$, $e^d_4=p^d_5-p^d_4$, and $e^d_5=p^d_5-p^d_1$; For Fig. \ref{directed}(c) (undirected graph case),  $e^d_k$ is set to $e^d_1=p^d_2-p^d_1$, $e^d_2=p^d_3-p^d_2$, $e^d_3=p^d_4-p^d_3$, $e^d_4=p^d_5-p^d_4$, and $e^d_5=p^d_1-p^d_5$. In the following, to show the validity and the advantages of the proposed algorithm, we will verify the asymmetric and global performance behaviors just by choosing different design parameters under a fixed controller. To simplify notations, we choose the identical performance function for each edge, implying that each agent has the same communication/sensing capabilities, namely, 
	\begin{equation*}
	\mathcal{P}_k(\overline{\delta}_k\beta_{k}(t))=\mathcal{P}(\overline{\delta}\beta(t))~\textrm{and}~\mathcal{P}_k(-\underline{\delta}_k\beta_{k}(t))=\mathcal{P}(-\underline{\delta}\beta(t))
	\end{equation*}
	for $k=1,\dots,5$, where $\beta(t)=(\beta_{0}-\beta_{f})\exp(-\lambda t)+\beta_{f}$.
	
	\begin{figure*}[t!]
		\centering	
		\includegraphics[width=6.7in]{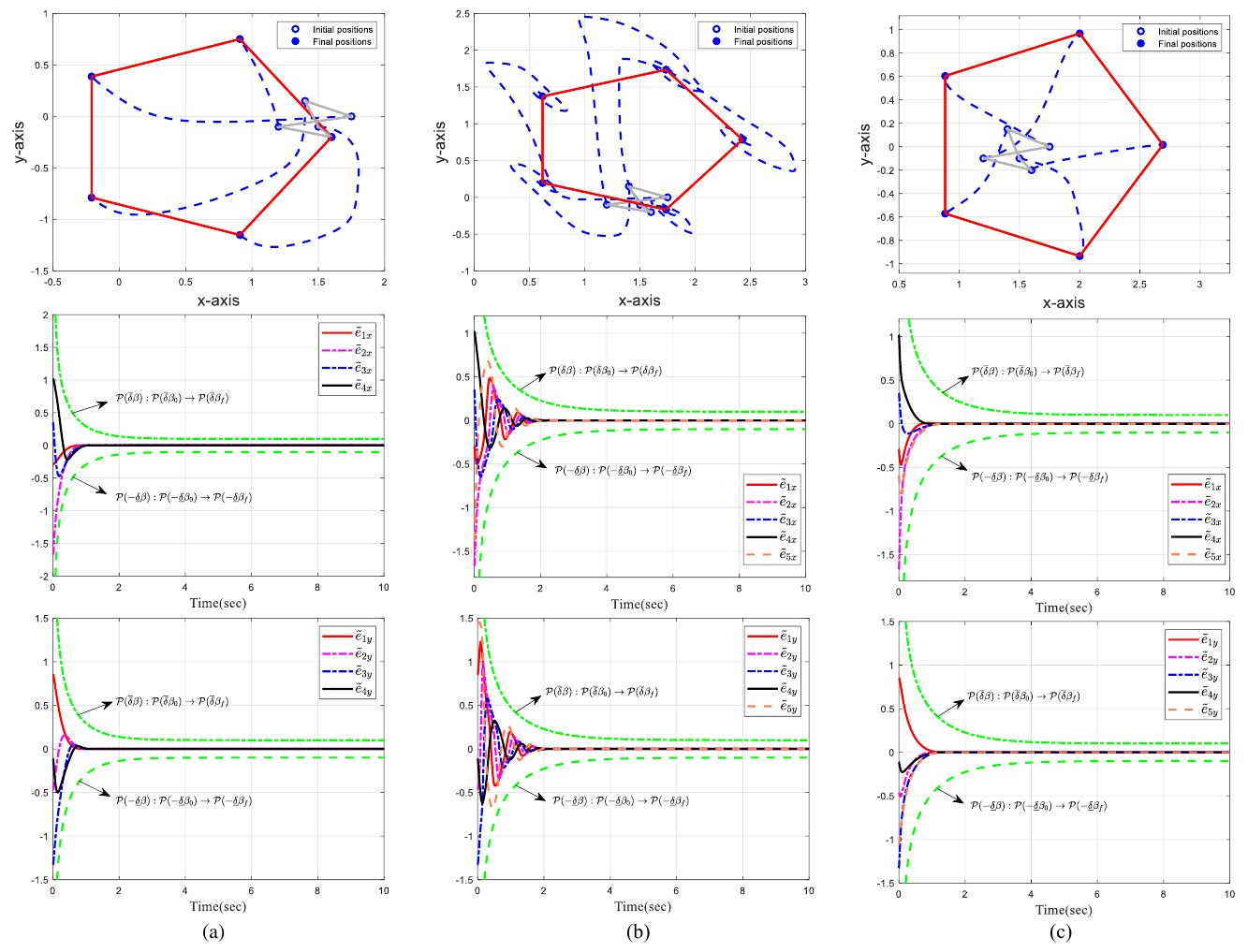}
		\caption{Global performance behavior. (a) directed spanning tree. (b) directed cycle, and (c) undirected graph cases.}\label{globalcase}
	\end{figure*}
	
	\subsection{Asymmetric Performance Behavior}\label{asymmetric}
	To verify that the proposed control is able to achieve the asymmetric prescribed tracking performance. In this part, for the directed spanning tree case, the agents' initial positions are set as $x_{1,1}(0)=[1,1.2]^\top$, $x_{2,1}(0)=[0.6, 2.3]^\top$, $x_{3,1}(0)=[-0.8, 2.1]^\top$, $x_{4,1}(0)=[-1.4, 0.5]^\top$, and $x_{5,1}(0)=[-0.2,0.35]^\top$. For the directed cycle and undirected graph case, the agents' initial positions are set as $x_{1,1}(0)=[1.2,2]^\top$, $x_{2,1}(0)=[0.5, 3]^\top$, $x_{3,1}(0)=[-0.8, 2.5]^\top$, $x_{4,1}(0)=[-0.9, 1.5]^\top$, and $x_{5,1}(0)=[0.4,1]^\top$; the initial velocities for each agent are set as $v_i(0)=\textrm{zeros}(2,1)$. To fulfill the initial constraint, the key design parameters $\underline{\delta}$, $\overline{\delta}$ and $\beta_{0}$ are selected as $\underline{\delta}=0.3$ and $\overline{\delta}=\beta_{0}=0.8$ such that each initial relative error falls within the initial prescribed boundary, i.e., $-0.25=\mathcal{P}(-\underline{\delta}\beta_{0})<\tilde{e}_{kq}(0)<\mathcal{P}(\overline{\delta}\beta_{0})=0.83$, $q\in \{x,y\}$ and $k=1,\dots,5$. The other parameters are chosen as: $\lambda=1$, $c_1=1$, $c_2=5$, $\Gamma_i=I_2$, and $\beta_{f}=0.1$. By implementing adaptive control law (\ref{u}) and (\ref{adaptive}), the simulation results are shown in Fig. \ref{asymmetrycase}. From Fig. \ref{asymmetrycase}(top) it is observed that the five agents eventually achieve and maintain the target formation shape during their motions. The responses of edge errors are depicted in the middle and bottom of Fig. \ref{asymmetrycase}, respectively, from which it can be seen that in all cases—(a) directed spanning tree, (b) directed cycle, and (c) undirected graph—the edge errors asymptotically converge to zero within the prescribed performance bounds over time.
	
	It should be noted that in order to guarantee the asymmetric performance behavior of the relative errors, it is crucial to carefully choose the key design parameters $\underline{\delta}$, $\overline{\delta}$ and $\beta_{0}$ to meet the initial constraints (i.e., $\mathcal{P}(-\underline{\delta}\beta_{0})<\tilde{e}_{kq}(0)<\mathcal{P}(\overline{\delta}\beta_{0})$). 
	This situation also exists in \cite{bechlioulis2016decentralized,verginis2017robust,cheng2022fixed,wang2015prescribed,karayiannidis2012multi,macellari2016multi, bechlioulis2014robust,chen2020leader,stamouli2022robust,huang2024prescribed}. However, this parameter selection process is exceedingly tedious and time-consuming, for example, in an $N$-agent MAS with $m$ edges, it necessitates the design and execution of a performance boundary selection algorithm at least $m$ times to identify feasible parameters. To circumvent such challenges, this work proposes a solution by considering global performance characteristics, as discussed below.

	\subsection{Global Performance Behavior}
	
	In this part, the agents' initial positions are set as $x_{1,1}(0)=[1.6,-0.2]^\top$, $x_{2,1}(0)=[1.2, -0.1]^\top$, $x_{3,1}(0)=[1.75, 0]^\top$, $x_{4,1}(0)=[1.4, 0.15]^\top$, and  $x_{5,1}(0)=[1.5,-0.1]^\top$, and the initial velocities are also set as $v_i(0)=\textrm{zeros}(2,1)$. The key design parameters $\underline{\delta}$, $\overline{\delta}$, and $\beta_{0}$ are selected as $\underline{\delta}=\overline{\delta}=\beta_{0}=1$, i.e., $-\infty=\mathcal{P}(-\underline{\delta}\beta_{0})<\tilde{e}_{kq}(0)<\mathcal{P}(\overline{\delta}\beta_{0})=\infty$, $q\in \{x,y\}$, which shows that there is no any constraint on the initial states. The other parameters are chosen as: $\lambda=1$, $c_1=2$, $c_2=5$, $\Gamma_i=I_2$, and $\beta_{f}=0.1$. By implementing adaptive control law (\ref{u}) and (\ref{adaptive}), the simulation results are shown in Fig. \ref{globalcase}. From Fig. \ref{globalcase}(top) it can be seen that the five agents eventually achieve and maintain the target formation shape during their motions. Figs. \ref{globalcase} middle and bottom show that the edge errors asymptotically converge to zero within the prescribed boundary, which indicates that the performance behaviors are independent of initial conditions, illustrating the effectiveness of the proposed approach.

	\section{Conclusion}\label{VI}
	In this paper, a novel adaptive formation control algorithm has been proposed for a group of agents described by uncertain high-order nonlinear dynamic models. The features that distinguish our method from the existing ones include: 1) the relative error constraints between the agents with physical meaning are taken into account, which makes the developed algorithm advantageous for practical applications;  2) it is able to fulfill various performance requirements merely by selecting design parameters a priori, making the control redesign and stability reanalysis not required; and 3) the dynamic model and underlying graph considered in this work are more general such that most of existing results are the special cases of the proposed control. The simulation results have shown the efficacy of our approach. Future research may consider multiple time-varying performance constraints and derive a formation control algorithm in the local reference frame.

\bibliographystyle{IEEEtran}
\bibliography{reference}

\end{document}